\documentclass[sigconf,nonacm]{acmart}

\AtBeginDocument{%
  }

\copyrightyear{2022} 
\acmYear{2022} 
\setcopyright{acmcopyright}\acmConference[WebSci '22]{14th ACM Web Science Conference 2022}{June 26--29, 2022}{Barcelona, Spain}
\acmBooktitle{14th ACM Web Science Conference 2022 (WebSci '22), June 26--29, 2022, Barcelona, Spain}
\acmPrice{15.00}
\acmDOI{10.1145/3501247.3531582}
\acmISBN{978-1-4503-9191-7/22/06}

\begin{document}

\title{You Have Earned a Trophy: \\ Characterize In-Game Achievements and Their Completions}

\author{[Please cite the WebSci'22 version of this paper]\\Haewoon Kwak}
\email{haewoon@acm.org}
\orcid{0000-0003-1418-0834
}
\affiliation{%
  \institution{Singapore Management University}
  \country{Singapore}
}

\renewcommand{\shortauthors}{Kwak}

\begin{abstract}
    Achievement systems have been actively adopted in gaming platforms to maintain players' interests. Among them, trophies in PlayStation games are one of the most successful achievement systems. While the importance of trophy design has been casually discussed in many game developers' forums, there has been no systematic study of the historical dataset of trophies yet. In this work, we construct a complete dataset of PlayStation games and their trophies and investigate them from both the developers' and players' perspectives. 
\end{abstract}



\maketitle

\section{Introduction}

How to make gamers not leave but keep playing a game is the  million-dollar question in the gaming industry~\cite{park2017achievement}. 
That is relevant not only for online games but also for video games. 
A variety of approaches has been proposed and used in practice, from in-game elements to social factors.  
For example, as in-game elements, developers sometimes include  collectible items, prepare challenging quests that take time to clear, or add minigames to compete for higher scores. As social factors, many gaming platforms support players to make friends with each other and allow sharing content, such as streams of users' playing games, screen captures, or progress with their friends. 

An achievement system is one of these approaches~\cite{jakobsson2011achievement}, which is sometimes called a  trophy~\cite{ps_trophy} or badge~\cite{easley2016incentives}. 
The achievement is given to players when they satisfy a predefined condition. 
The range of conditions is diverse, from starting a game to killing enemies, winning matches, or collecting items. 
Modern games use these achievements to promote diverse gameplay and give positive feedback by setting unusual play or difficult challenges as the achievement condition~\cite{przybylski2010motivational}.
In other words, achievement systems can work as an additional layer of gameplay in traditional games, giving positive feedback to users. 
Since it is well known that feedback is crucial in interactive systems~\cite{sundar2012social},  achievement systems play a vital role in maintaining the interest of players in modern games and consequently, they have been widely adopted in diverse gaming platforms, including PlayStation, Xbox, and Steam. 
Thus, it is not surprising that the achievement system has attracted some attention too from academics since it was first  introduced~\cite{siu2016reward,hamari2011framework,lewis2013features,cruz2017need}. 
Most researchers in the field to date have tended to conduct user interviews to understand the player perception of achievement systems. Unfortunately, to the best of our knowledge, there has been no study yet to reveal a platform-wide comprehensive picture of achievements and their interaction with users, probably due to a lack of data.

To fill this gap, we aim to collect large-scale data on achievements and explore it in a data-driven way. Our dataset includes all the achievements (trophies) of a complete set of games released on Sony Interactive Entertainment's platforms, including PlayStation 3, 4, 5, Vita, and VR, as well as their completion rates (i.e., how many players  successfully got a certain trophy). 
While multiple platforms offer similar achievement systems, we focus on the PlayStation platforms in this work because i) PlayStation consoles are the biggest video console platforms equipped with an achievement system across all the console generations, and ii) many game studios tend to release their games on multiple platforms, including on the PlayStation consoles. 

We examine the collected data from both the developers' and players' perspectives to understand the design trend of trophies and their completion rates, respectively. 
Although more than a decade has passed since the  achievement system was first introduced in the gaming industry and game designers have been sharing their experience~\cite{gridsagegames2018,gdc2013designing}, it is still unclear what the trend is in this area and how challenging the achievements are to gamers. 
Our work aims to reveal the answers to these questions in a data-driven way.

From the developers' perspective, we begin by analyzing how developers have adapted to trophy systems over time, particularly regarding the number of trophies they include per game. 
We then look into this aggregated ``number'' of trophies per game in detail. We dissect the number of trophies included per game based on their levels, which are Bronze, Silver, Gold, and Platinum, and examine how developers set the  different levels of trophies per game. We find that the number of trophies of the lowest level (i.e., Bronze) per game has decreased over time, but that of the higher one (i.e., Gold) has  increased. This indicates that there is a trend for games to create more challenging trophies over time. 
To further analyze how developers design the condition for when a certain trophy is awarded, we apply semantic role labeling and part-of-speech tagging to the trophy condition descriptions to extract the ``who did what to whom''~\cite{zhou2015end}. Then, we apply SentenceBERT~\cite{reimers-gurevych-2019-sentence} to cluster the trophy conditions. 

From the players' perspective, we focus on how many people actually get trophies by achieving the predefined conditions. Our findings reveal that this changes over time and varies with playtime, genre, and repetitiveness in the trophy conditions. 

Last, we conduct a preliminary analysis of the  cross-platform games that have been released on PlayStation and Xbox platforms. We test the generalizability of our findings between the platforms. 

Our contributions are as follows:

\begin{itemize}
    \item We collect large-scale data on the PlayStation trophies in all the available games since the trophy system was first  adopted in 2007. By comparing the statistics computed from all PlayStation gamers and a gamer community, we demonstrate a potential bias of the gamer community for studying in-game achievement systems. 
    \item We examine an interplay between developers and players regarding the trophies\textemdash how to design and achieve them. Through a series of quantitative analyses, we reveal the temporal trends of trophy design and completion. These will provide valuable insights for developers. 
\end{itemize}

The rest of this paper is organized as follows: Section 2 introduces the background of this study from the perspective of players' motivation and in-game achievements. Section 3 explains our data collection. Section 4 outlines analyses from two perspectives. Section 5 shows our analyses and results from the developers' perspective. Section 6 presents those from the players' perspective. Section 7 shows a preliminary analysis of cross-platform games. In Section 8, we summarize our findings and discuss the limitations and future directions.

\section{Background}

The self-determination theory~\cite{deci2000and} provides a foundation for understanding human motivation based on social context to satisfy universal human needs and values.  
According to the theory, we can understand why players may engage in games because (1) games are fun and players just wish to do so (intrinsic motivation), and (2) there are rewards or punishments to encourage them to play (extrinsic motivation). 
Several studies suggest that external rewards might be effective in the short term but fail to increase intrinsic motivation~\cite{deci1999meta}. 
In particular, when such rewards become unavailable, players lose their motivation to keep doing a task~\cite{lepper1973undermining}.

While it is still debatable, it can be said that in-game achievements offer both perspectives simultaneously~\cite{cruz2017need}. 
Cruz et al.~\cite{cruz2017need} conducted a focus group study with 36 people who owned a video game console and had played $>=$ 50 hours of video games in the last six months. 
They observed mixed responses. Some players described in-game achievement systems as giving positive feedback, encouraging diverse ways to play games, and boosting their self-esteem. However, some other players felt that the  achievements were burdens because they represented extra tasks to complete.
In this work, we examine a platform-wide completion rate of achievements and its changes over time, which can imply how gamers overall have engaged with achievement systems.

 
Consalvo~\cite{consalvo2009cheating} proposed the notion of gaming capital, which is ``credit amassed within the gaming community on the basis of specialized knowledge such as how to unlock hidden badges.''
In that sense, game trophies can be considered as a tangible form of gaming capital~\cite{sotamaa2010achievement}. 
Gaining more trophies indicates a higher level of players' knowledge and experience. 
Since players' specialized experience is translated into trophies, users can compare their ranks with their peers, which provides a  social context.   

Hamari and Eranti~\cite{hamari2011framework} reviewed available achievement systems, interviewed developers and players, and conducted an observational study to build a framework for an achievement design. 
They proposed three components of an achievement: a signifier, a completion logic, and a reward. The signifier is the name, image, or description of an achievement, and the completion logic is a rule to specify when an achievement has to be awarded. In our context, a trophy description can be mapped into the signifier, a trophy condition into the completion logic, and a trophy awarded into the reward. 

One study on player profile data collected from a gamer community~\cite{wells2016mining} is the most relevant study to this work. 
It used a sample of 30,227 players' profiles (1.4\% of the registered users in the community) and analyzed their trophies from 3,212 games from the community. However, most of their experiments rely on basic statistics, such as computing an average or Pearson correlation coefficient. 
We provide a more comprehensive picture of game trophies with a consideration of the trophy levels, trophy conditions, genres, and play time by combining the data from multiple sources.  

\section{Datasets}

To study the trophies of PlayStation games in a data-driven way, we collect relevant data from two different sources: PSNProfiles~\cite{psnprofiles} and HowLongToBeat~\cite{howlongtobeat}. 

\subsection{PSNProfiles}
\label{subsec:psnprofiles}

PSNProfiles~\cite{psnprofiles} has maintained detailed information on all the PlayStation games available on the market since 2007 and their trophies. For games, it collects and shows developer, publisher, genre, release date, and more. For trophies, it collects the trophy's name, condition to earn, completion rate (proportion of users who got the trophy to users who own the game), and trophy level (e.g., Bronze, Silver, Gold, Platinum). PSNProfiles additionally compute the completion rate among its registered users only. 

We build a web crawler for collecting game and trophy information from PSNProfiles. To avoid any potential burden to the service, we set a one-minute interval between our requests and slowly collect the data. As a result, we obtain detailed information of 13,792 games on their 377,938 trophies. 

\subsection{HowLongToBeat}
\label{subsec:howlongtobeat}

HowLongToBeat (HLTB)~\cite{howlongtobeat} offers an interesting statistic that is not available on other services, which is the play time of a game. This statistic is based on gamers' self-reports. As there are large variations in individual's playtime, HLTB introduces a guideline that divides play style into three~\cite{howlongtobeat}: 

\begin{quote}
\begin{itemize}
    \item Main Story (Required): 
You complete the main objectives, just enough to see the credits roll.
    \item Main Story and Additional Quests/Medals/Unlockables: 
You take your time, discover and complete additional tasks not required.
    \item Completionist (100\%): 
You strive for every achievement, every medal and conquer all that the game has to offer.
\end{itemize} 
\end{quote}

Using HLTB's search interface, we collect games released for PlayStation platforms:  PlayStation 3, PlayStation 4, PlayStation 5, PlayStation Vita, and PlayStation VR. As the platforms before PS3 did not support trophies, they are out of the scope of this work. 
Similar to PSNProfiles, we slowly collect play time information of games at one-minute time  intervals. As a result, we collect data on 7,979 games' playtime. 
In contrast to PSN Profiles, HLTB does not have an entire catalog of released games. In that case, users can register games in HLTB's database as well as their play time. 


\section{Outline of the Analyses}

Our aim is to examine game trophies from two different angles. 
One is to understand \emph{how game developers have designed trophies}. In this regard, there are several constraints and considerations in designing trophies. 
In $\S$\ref{sec:developer}, we begin with the number of trophies per game, which is one of the most noticeable characteristics. We look into how many trophies are available per game and whether there are any temporal trends in the number of trophies in games. 
We then move on to the levels of the trophies. Not all trophies are equal in terms of their ``worthiness''. Developers assign one of four levels, which are Bronze, Silver, Gold, and Platinum, to each trophy. We analyze how many trophies are designed for each level and how this changes over time. 
Next, we dig into the conditions to earn trophies. To understand their semantics more systematically, we run semantic role labeling and cluster them based on their vector representations. Finally, we check how closely the trophies are connected to the game's plot through the hidden trophies included. 

The other angle is to understand \emph{how game players have interacted with trophies}. In $\S$6, a central question is this: how many players  actually earn the trophies? We then examine how this changes between different periods, levels of trophies, genres, playtime, and repetitiveness in the conditions.  

Finally, we conduct a preliminary analysis of cross-platform games released on PlayStation and Xbox platforms, and test the generalizability of our findings between the platforms. 

\section{Game Developers' perspectives}
\label{sec:developer}

\subsection{Two Types of PlayStation Games}

Before examining the number of trophies per game, we need to consider two different types of PlayStation games. One is a full-price game, which usually costs around \$59, and the other is a simpler game, which is sometimes offered for free with other membership (e.g., PlayStation Plus Free Games). As these two types of games are fundamentally different, they also need to be considered separately in the analysis. 

To systematically divide games into these two types, we focus on the trophy points that each game has. 
In the PlayStation trophy system, trophies can be one of four levels, and each level offers  different points: Bronze (15 points), Silver (30), Gold (90), and Platinum (180). 

\begin{figure}[h!]
  \includegraphics[width=\columnwidth]{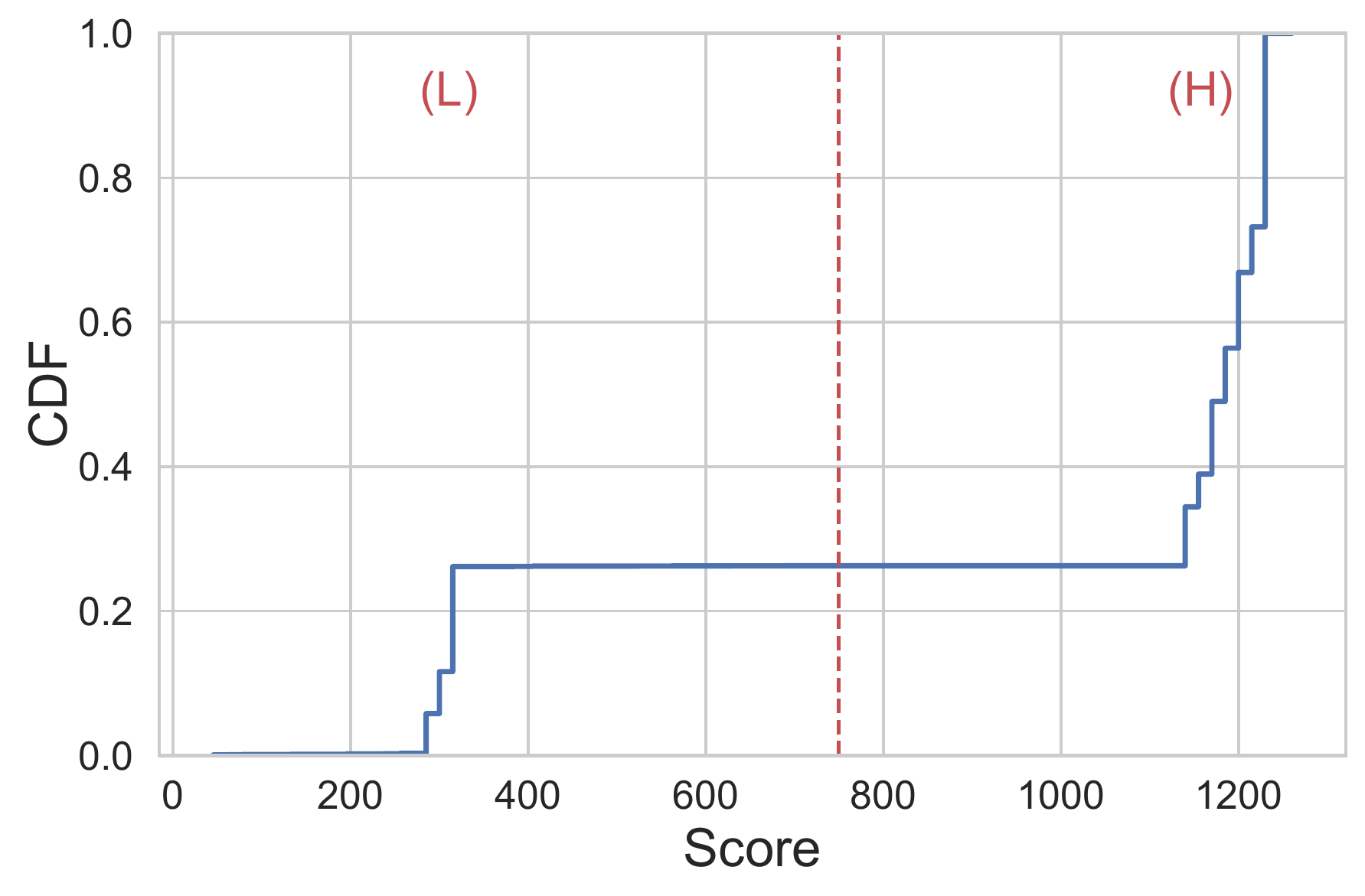}
  \caption{CDF of the total points of the trophies per game}
  \label{fig:CDF_score_per_game}
\end{figure}

This trophy point is summed across all the games at a user level. 
The aggregated points are shown in the user profiles in PlayStation Network, which is the online service for PlayStation gamers, making trophies meaningful beyond just an individual game and ensuring users are engaged in hunting trophies. 

Two types emerge if we sum the trophy points at an individual game level. 
Figure~\ref{fig:CDF_score_per_game} shows the  cumulative density function (CDF) of the summed scores. 
While they are not clear-cut, two clear peaks can be seen around 300 and 1,200. 
This suggests that some range of scores around 300 and 1,200 is allowed by Sony Interactive Entertainment, a platform holder, when designing  trophies for a single game. 
Based on this finding, we use 750 as a threshold and divide games into two categories: L (games whose summed trophy score is around 300) and H (games whose summed score is around 1,200). These divisions are also well matched with our intuition (L: relatively short game vs. H: typical full-price game), as shown in $\S$6. The primary focus of this work is type (H) games. 

Also, we would like to explain the recent change in the score of the Platinum trophy, as we briefly mentioned above. 
While the Platinum trophy's score changed to 300 from 180 in October 2020, the sum of trophy scores using 180 still stays at around 1,200. For example, a game released one year later has 1,230 scores using 180 instead of 300 for the Platinum trophy (see \url{https://psnprofiles.com/trophies/14481-shinrai-broken-beyond-despair}). 
Thus, we assume that the change does not affect trophy design at a game level but only impacts the user-level trophy score.

\subsection{Number of Trophies per Game}

\begin{figure}
  \includegraphics[width=\columnwidth]{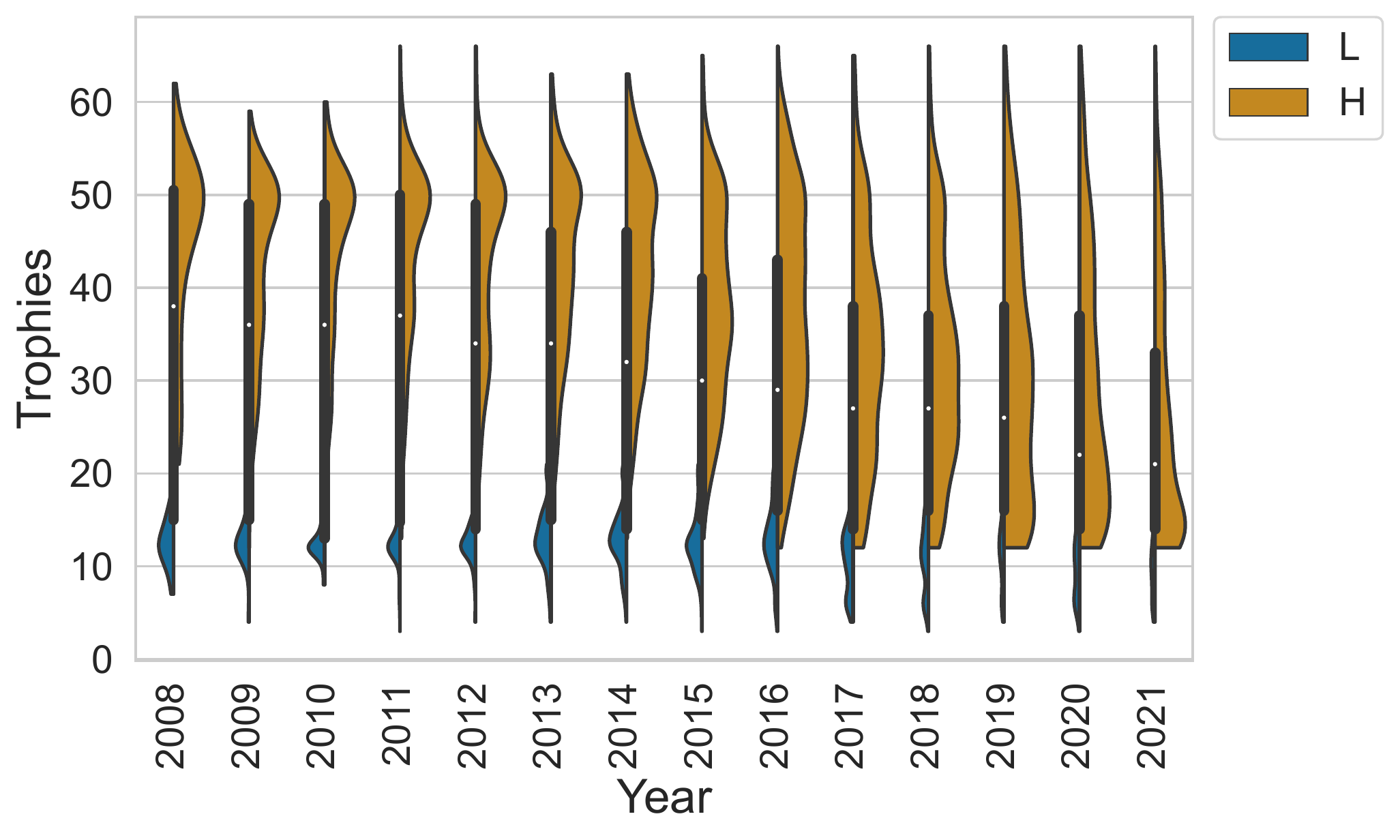}
  \caption{Number of trophies per game over time. L indicates games whose summed trophy score $<$ 750, and H indicates games whose summed score $>=$ 750.}
  \label{fig:number_of_trophies_per_game}
\end{figure}

We illustrate how many trophies are created per game in Figure~\ref{fig:number_of_trophies_per_game}.
The $x$-axis represents when the game was released on the market. 
The left half (L) of each `violin', which is colored blue, maps the games whose summed score is lower than 750, and the right half (H) of each `violin', which is colored golden brown, maps the games whose summed score is higher than or equal to 750. The division criteria between the two types is explained in $\S$5.1. As we mentioned earlier, type (L) games are  relatively short games, and (H) games are typical full-price games. 

We can see a clear trend whereby the number of trophies decreases over time. When the trophy system was newly introduced in PlayStation games\footnote{As only 7 games were published in 2007, we removed them from the temporal analyses.}, the mode (The most frequent number) of trophies of typical full-price games (H) reaches even at 51. 
The peak decreases over time, and in 2021, the mode was only 13. 
The percentage of games with more than 35 trophies was 77.4\% in 2008, but it quickly decreased and reached 25.2\% in 2021. 
The trend is weaker among relatively short games (L), but the median decreased from 12 in 2008 to 10 in 2021 ($p<0.05$ by t-test). 
This finding shows how developers have adapted to the new trophy system, where the direction of adaptation is to place lesser burdens on players. This could be a response to some players who feel that achievements are extra tasks to complete~\cite{cruz2017need}.
The outcome of this trend is also apparent in how the completion rate of trophies changes over time in $\S$6. 

\subsection{Levels of Trophies}
\label{subsec:levels_of_trophies}

\begin{figure}[h!]
  \includegraphics[width=\columnwidth]{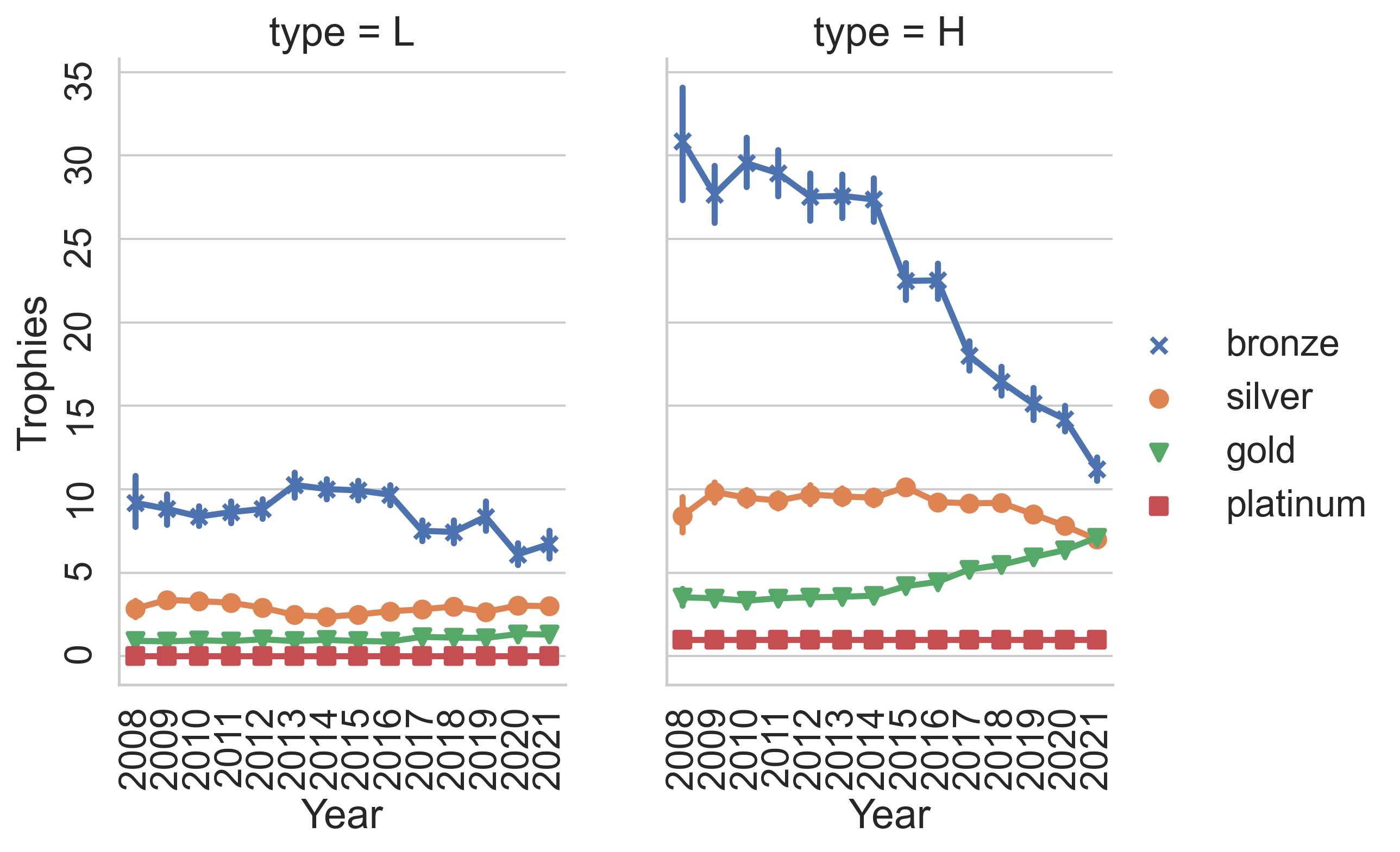}
  \caption{Number of trophies at each level per game over time}
  \label{fig:category_trophy_over_time}
\end{figure}

As we briefly explained in $\S$5.1, there are four levels of trophies. 
Figure~\ref{fig:category_trophy_over_time} shows how the number of trophies of different levels per game has changed over time according to the  game type (L or H). First, we see a strict constraint regarding the Platinum trophy: zero  for type (L) games, and only one for type (H) games. 
The Platinum trophy is the rarest and hardest to achieve. 
Second, there are interesting temporal trends among type (H) games: Bronze and Silver trophies decrease, but Gold trophies increase over time. 
Along with Figure~\ref{fig:number_of_trophies_per_game}, this finding shows that Bronze and Silver trophies, which are designed for more straightforward tasks, have become less common. 
By contrast, Gold trophies, which are designed for harder tasks, have been steadily increasing over time. As one Gold trophy (90 points) is worth as much as six Bronze trophies (15), a large cut of Bronze trophies is unavoidable if designers want to increase the number of Gold trophies while maintaining the same total score. 
This implies that the entire trophy design scheme has been changing; achievements for harder tasks are more frequently given, while those for simpler tasks are decreasing. This change is potentially an excellent way to give positive feedback to players for harder tasks and can lead to increase playtime, which game developers want the most. 

\subsection{Trophy Conditions}

A trophy is given to users when a corresponding condition is satisfied. Some conditions are trivial without needing additional effort (e.g., Bronze: Start the game), but some are repetitive and require lots of effort (e.g., Silver: Win 10,000 battles). 

To understand the conditions of trophies on a  large scale, we use semantic role labeling (SRL)~\cite{he2017deep}. An end-to-end approach for SRL based on deep neural networks has successfully discovered ``the predicate-argument structure'' of a sentence, which is who did what to whom~\cite{zhou2015end}. 

We apply the AllenNLP library~\cite{Gardner2017AllenNLP} to run our SRL analysis for the trophy conditions, which is a BERT-based SRL implementation~\cite{shi2019simple}. For example, when `collect 100 Death cards in the game' is a trophy condition, SRL can find `collect' as Verb\footnote{We use `Verb' to refer to a verb identified by SRL.} and `100 Death cards in the game' as Arg1 (theme or object argument). We then apply part-of-speech tagging to Arg1 and find the root noun, which is `cards' in the example. We lemmatize Verb and Arg1, which are `collect' and `card' in the example.  

\begin{table}
\centering \small
\caption{Top 10 most frequently appearing Verbs and top 5 Arg1s most frequently associated with those Verbs across all genres.}\label{tab:verb_arg1_all_genres}
\begin{tabular}{ll}
\toprule
Verb & Root noun of Arg1 \\
\midrule
complete & level, mission, game, chapter, quest \\
get & trophy, star, score, kill, medal \\
defeat & enemy, boss, monster, opponent, king \\
win & match, game, race, battle, event \\
collect & trophy, coin, item, star, treasure \\
kill & enemy, boss, player, monster, zombie \\
find & word, secret, treasure, collectible, item \\
have & level, ball, conversation, gold, coin \\
finish & game, level, chapter, mission, race \\
use & attack, skill, ability, player, weapon \\
\bottomrule
\end{tabular}

\end{table}

Table~\ref{tab:verb_arg1_all_genres} shows the top 10 Verbs and top 5 Arg1s that are the most frequently associated with each Verb. A pair of the Verb and the root noun of Arg1 presents what kinds of conditions are frequently set by developers. 
A `complete something'-type condition, which is called `Completion' in \cite{montola2009applying}, is the most frequent. It can be applied for a level, mission, game, chapter, or quest. Sometimes `finish' can be used instead of `complete' for referring to  similar types of trophy conditions. 
The second-most-common type of condition is `get something', which is `Collection' in \cite{montola2009applying}. This is usually for  in-game collectible items (e.g., coins, stars, treasures, or medals.) or scores. We find that `collect' and `find' can be used instead of `get'. 
The third-most-common type of condition is `defeat something'. An enemy, boss, monster, opponent, or king may be an object of this condition. This condition is also related to `win' and `kill' conditions. While this type is not explicitly mentioned in \cite{montola2009applying}, its commonness  suggests that the `defeat something'-type condition should get enough attention as a separate condition rather than a part of a `complete something'-type condition. 

\begin{table}
\centering \small
\caption{Uniquely appearing Verbs (in the top 10) from the top 5 popular  genres}\label{tab:verb_arg1_each_genre}
\begin{tabular}{lll}
\toprule
Genre & Verb & Root noun of Arg1 \\
\midrule
Adventure & obtain & trophy, weapon, power, job, ability \\
Arcade & beat & level, boss, game, stage, score \\
& destroy & enemy, boss, tank, block, car \\
Sport & award & title, acclaim, style \\
& score & goal, try, point, kick, touchdown \\
& play & match, game, total, round, season \\
& earn & trophy, medal, point, star, total \\
& perform & trick, move, total, tackle, combo \\
& hit & run, ball, opponent, double, slam \\
Role-playing (RPG) & obtain & trophy, affinity, item, soul, weapon \\
& be & you, we, level, rate, number \\
& reach & level, affinity, chapter, end, circle \\
& acquire & trophy, skill, gold, treasure, point \\
Shooter & destroy & robot, enemy, vehicle, object, drone \\
\bottomrule
\end{tabular}
\end{table}

Some Verbs frequently appear more in one genre than the others, meaning that developers put effort into creating trophies that match with genre-specific narratives.
Table~\ref{tab:verb_arg1_each_genre} presents unique Verbs for each genre by comparing the top 10 Verbs in each genre with those across all genres, showing the unique tastes of each genre; 
Arcade or Shooter games use the `destroy' Verb (e.g., destroy tanks or robots).  Adventure and Role-playing (RPG) games use the `obtain' Verb to put more emphasis on the planned efforts to get something (from Cambridge Advanced Learner's Dictionary \& Thesaurus),  
while in Sport games, 6 out of the 10 most appearing Verbs are unique, demonstrating their  distinctive game mechanics compared to other genres. 

\begin{figure}[h!]
  \includegraphics[width=70mm]{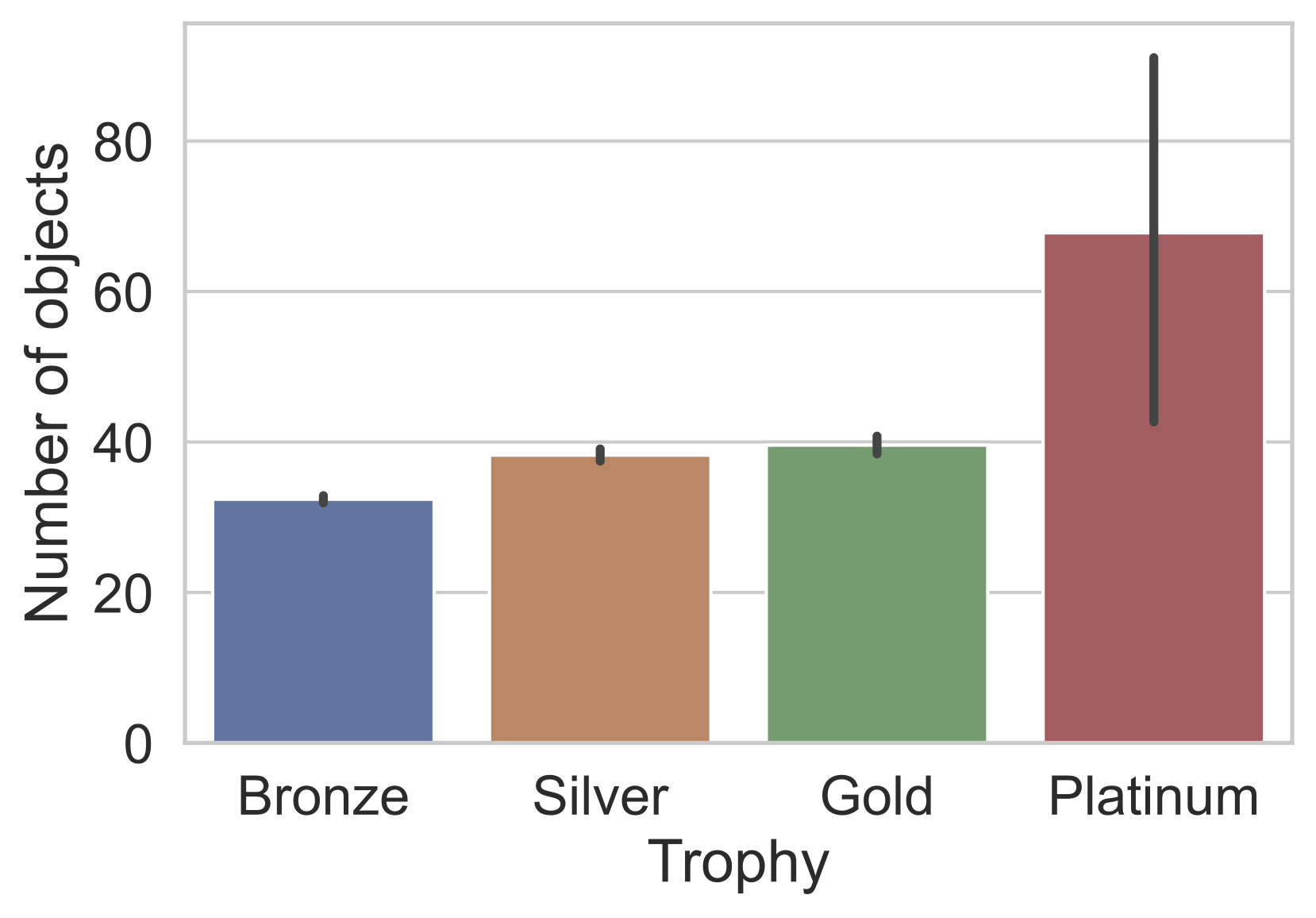}
  \caption{Number of objects per trophy category}
  \label{fig:N_trophy_cat}
\end{figure}

From the trophy condition, we additionally extract the number of objects to be satisfied. For example, 10 from `kill 10 enemies' and 100 from `Collect 100 seeds' become the number of objects.
Figure~\ref{fig:N_trophy_cat} shows the numerical quantifiers included in the conditions for the different levels of trophies. 
We can see that these increase with the level of the trophies, showing the increasing difficulties of achieving the trophies. Also, as `all' and `every' are not counted as numerical quantifiers but are commonly used expressions in trophy conditions, we compute the percentage of expressions that include `all' or `every' in the  trophy conditions per level. We find that 70.3\% of the Platinum trophies conditions have `all' or `every' in them, which suggests that the Platinum trophy is likely to be the final trophy given to players as an award for  completion. While the percentage is not as high as for Platinum, other levels of trophies also have `all' or `every' in their conditions (6.8\%, 13.6\%, and 15.7\% for Bronze, Silver, and Gold, respectively). This suggests that game developers tend to set more challenging trophies in Silver and Gold than Bronze.

\begin{figure}[h!]
  \includegraphics[width=\columnwidth]{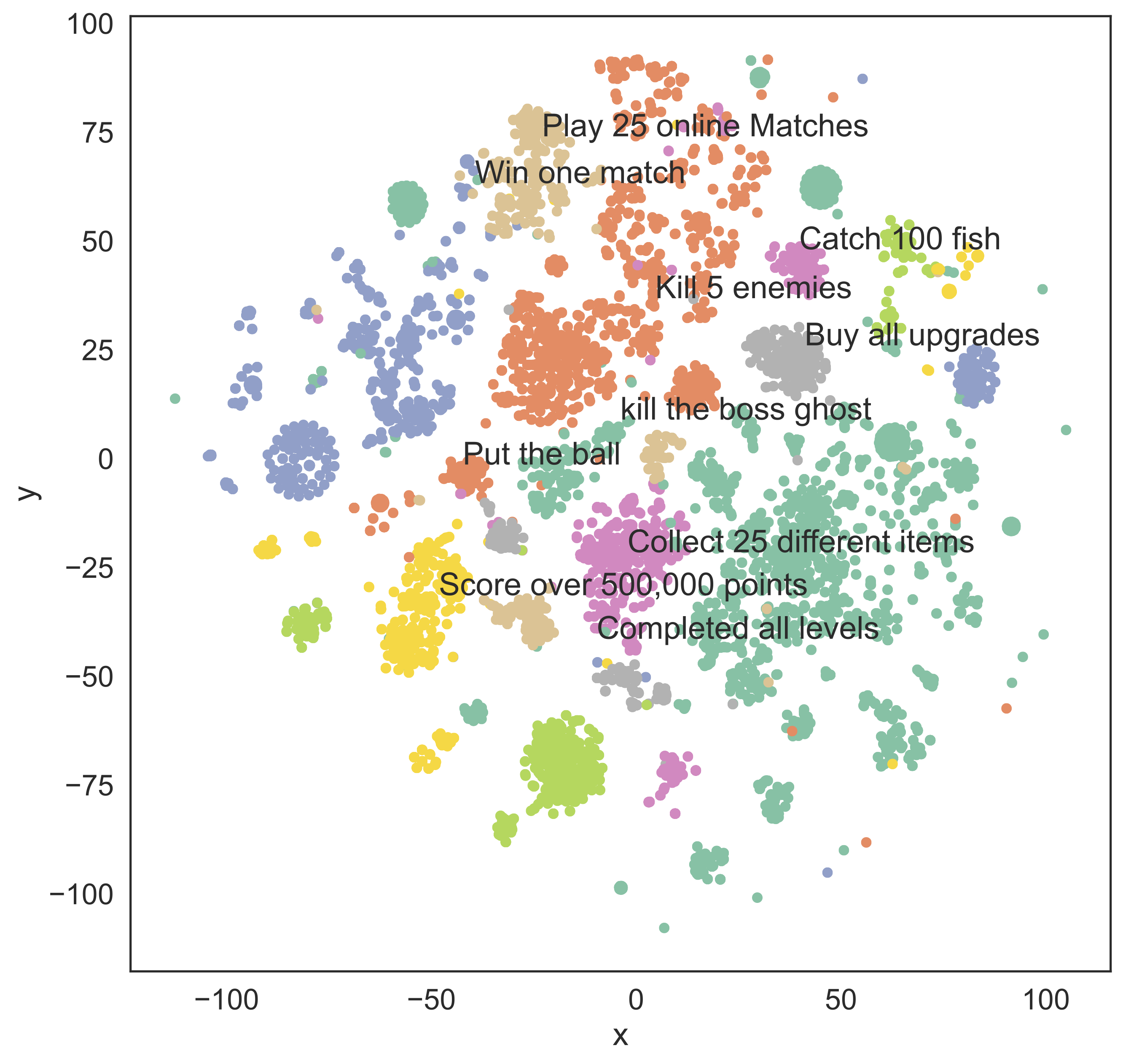}
  \caption{t-SNE visualization of trophy conditions clusters}
  \label{fig:tsne_trophy}
\end{figure}

As an extension of the semantic role labeling, we cluster the trophy conditions based on their vector representations. We use SentenceBERT~\cite{reimers-gurevych-2019-sentence} to get the representations of a trophy condition and then cluster them based on cosine similarity between the representations. 
We set the similarity threshold as 0.75 and run the fast clustering algorithm in Sentence Transformer library, which supports large-scale data~\cite{reimers-gurevych-2019-sentence}. We then show a t-SNE visualization of the identified clusters in Figure~\ref{fig:tsne_trophy}. For presentation purposes, we show clusters whose size (number of trophies) is bigger than 100 and the trophy conditions for the top 10 clusters.  
The results are well aligned with Tables~\ref{tab:verb_arg1_all_genres} and \ref{tab:verb_arg1_each_genre}. The Verbs of `complete', `collect', and `win' appear at the center of some clusters. Some conditions, such as `Buy all upgrades', do not specifically include top 10 Verbs but are related to them or include genre-specific Verbs; for example, `catch' and  `buy' are related to `collect', `play' is one of the top 10 verbs in Sport games, and so on. Our clusters again well align with the taxonomy manually built in~\cite{montola2009applying}. If we were to extend this approach of clustering representations by adjusting the thresholds, we would get a comprehensive taxonomy of trophy conditions while minimizing subjective human decisions in building the taxonomy, which is considered a hurdle in the replication of previous studies~\cite{lewis2013features,wells2016mining}.

\subsection{Hidden Trophies}

\begin{figure}[h!]
  \includegraphics[width=\columnwidth]{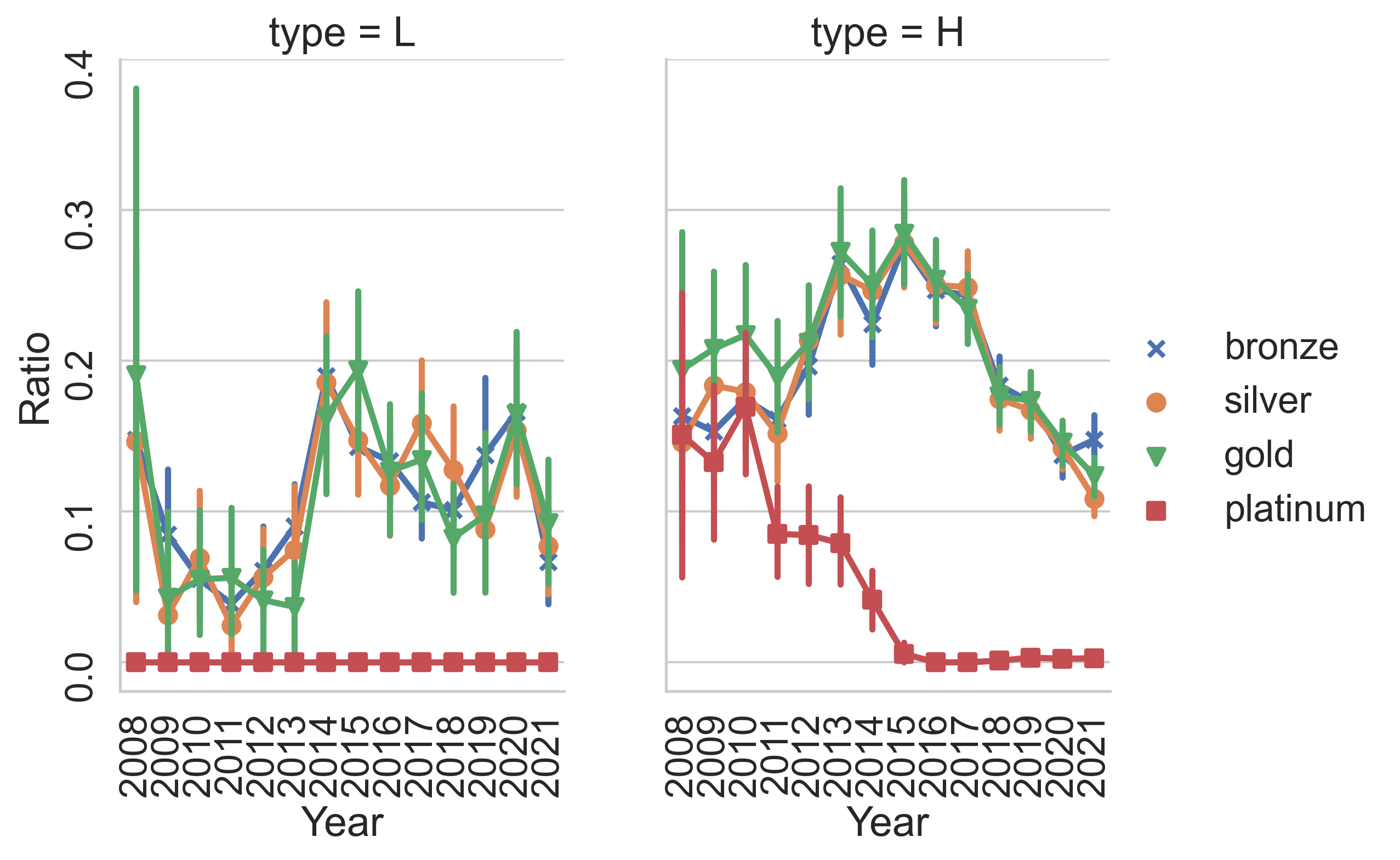}
  \caption{Ratios of hidden trophies of each category per game over time}
  \label{fig:secret_trophy_over_time}
\end{figure}

While a trophy condition is typically visible at default to players so that they can try to achieve it, some trophies (called `hidden' trophies) make their conditions invisible. 
They become visible only when the condition is satisfied and the player gets the trophy. One of the reasons behind having hidden trophies is that they are closely related to the game's plot. In that case, revealing the conditions of such trophies could act as a spoiler pre-revealing some aspect of the story, and thus, developers keep them hidden. 

Figure~\ref{fig:secret_trophy_over_time} presents how the ratio of hidden trophies per game changes over time. Type (H) games show an  increasing trend of the ratio of all the hidden trophies, except for Platinum initially, with  peaks visible at 2013 and 2015. At that time, more than one out of every four trophies ($>$0.25) were hidden. 
The ratios, however, quickly decreased after that. In 2021, the ratios stayed between 0.10 and 0.15. 
The ratio of hidden Platinum trophies was also highest in the early years. In 2010, more than one out of every six Platinum trophies were hidden (0.17). This ratio also rapidly decreased  over time and stayed at almost zero in 2021. 
This finding, similar with Figure~\ref{fig:number_of_trophies_per_game}, demonstrates how developers adapt to hidden trophies over time. The developers start to create fewer trophies related to the plot of the game and start making more trophies visible. These changes allow players to target trophies and to share trophy information without needing to worry about potential spoilers. As a result, players can be more engaged to achieve the trophies. 

\section{Players' Perspectives}

Now we look into trophy data from the players' perspectives.

\subsection{Completion Rates of the Trophies}

In $\S$\ref{subsec:psnprofiles}, we mention that PSNProfiles shows the completion rate of the  trophies, which is the percentage of users who got the trophies to the users who own the game, using two sources: 1) the entire PlayStation Network (PSN) players as computed by Sony Interactive Entertainment and 2) the registered users in PSNProfiles. 

\begin{figure}[h!]
  \includegraphics[width=\columnwidth]{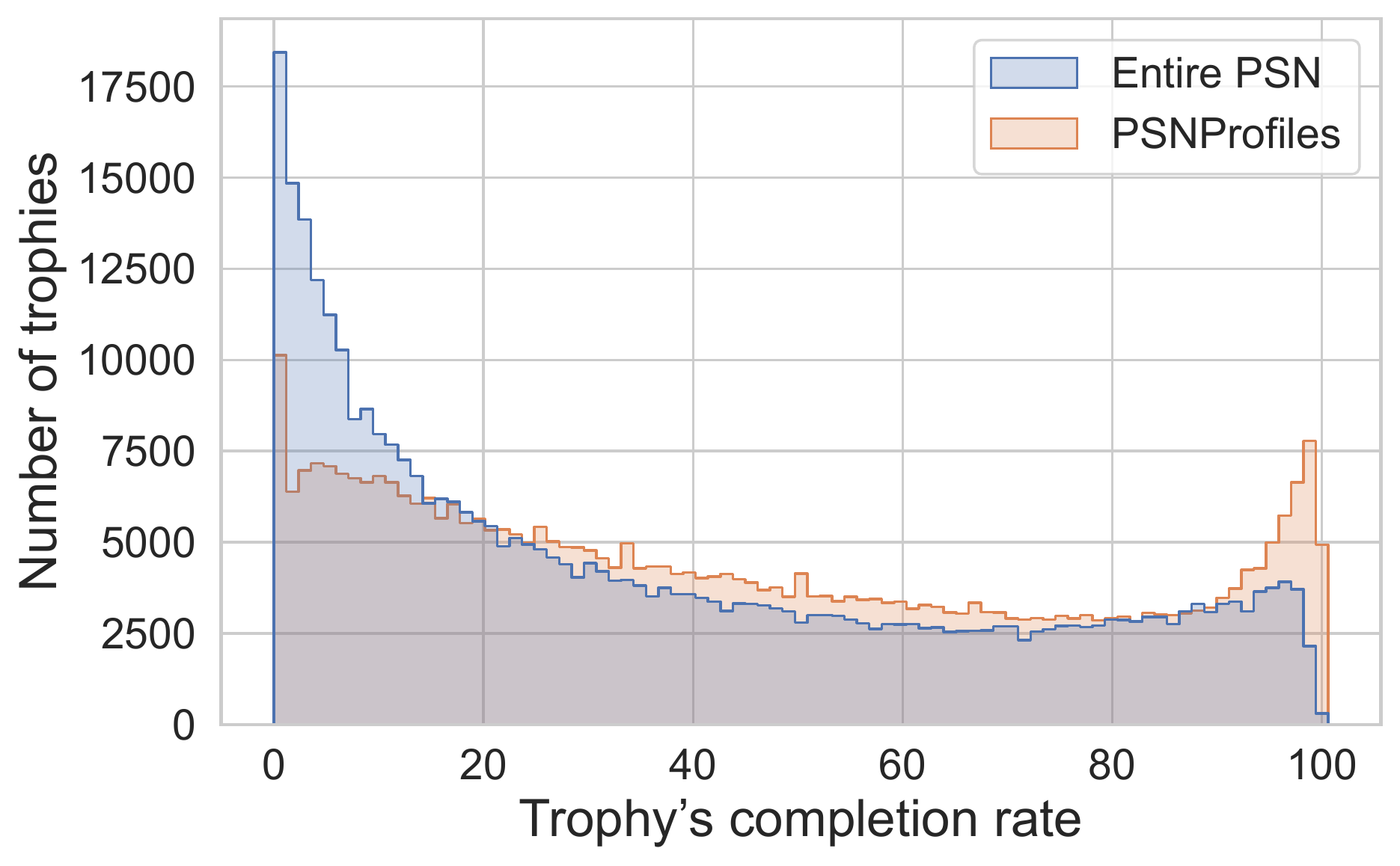}
  \caption{Completion rate of the trophies computed from two different sources}
  \label{fig:trophy_rarity}
\end{figure}

Figure~\ref{fig:trophy_rarity} shows the completion rate of trophies from both sources. The high peak of the trophies with a completion rate of 100 clearly shows that the registered users in PSNProfiles are highly engaged in seeking trophies.
This is not so surprising given that  PSNProfiles is a dedicated online community for PlayStation game trophies, and it is only natural that its users should be more active in chasing trophies. Thus, we reaffirm that PSNProfiles data should be carefully analyzed and generalized as \cite{wells2016mining} conjectured. 
Also, the figure demonstrates why it is hard to study individual-level trophy records; game-focused communities, which are likely to be a source of individual-level trophy data, are skewed toward expert gamers compared to an entire population. In the rest of this paper, we focus on the completion rate of the entire PSN only. 

\begin{figure}[h!]
  \includegraphics[width=\columnwidth]{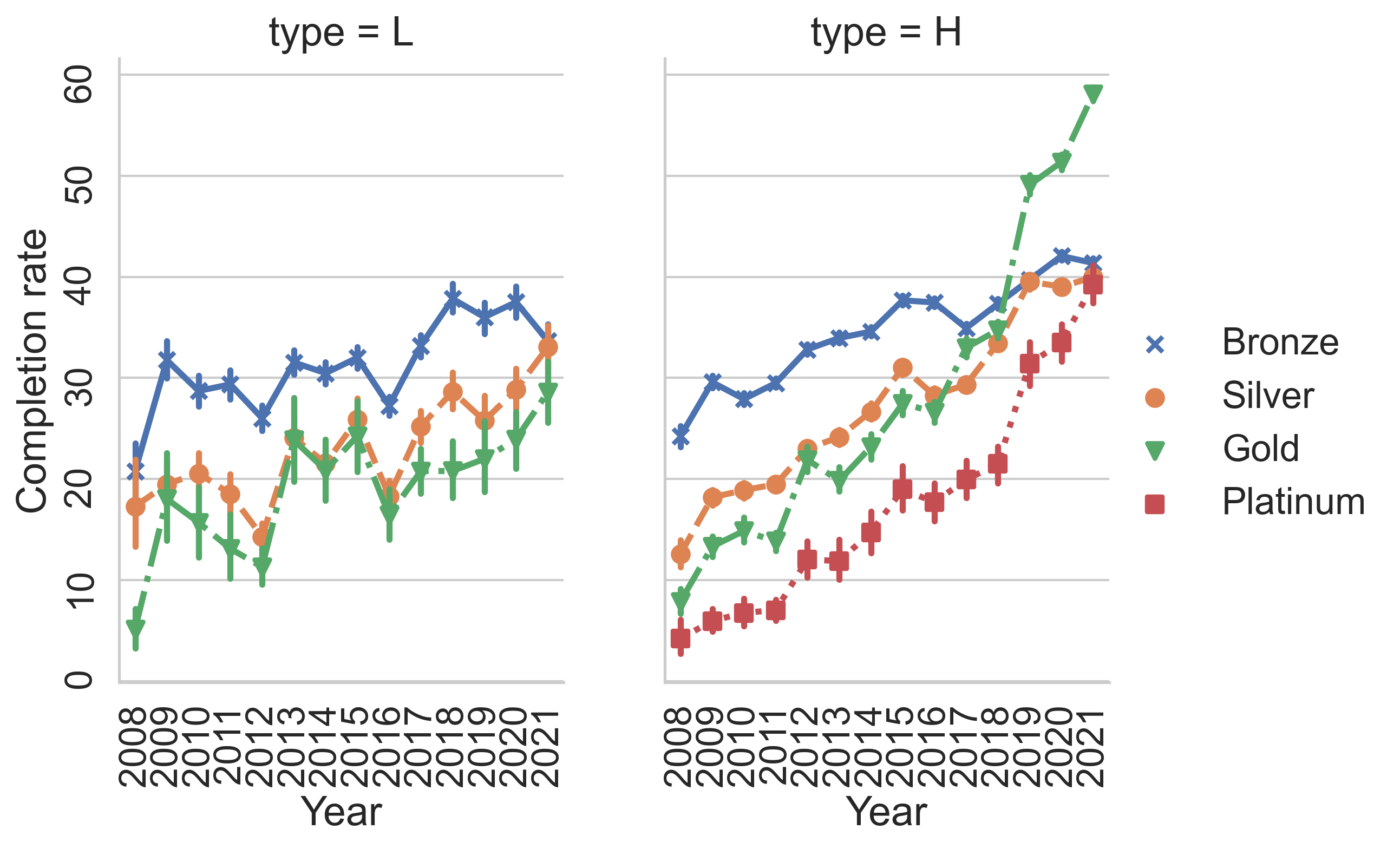}
  \caption{Completion rate of trophies per category over time}
  \label{fig:trophy_rarity_per_category}
\end{figure}

Figure~\ref{fig:trophy_rarity_per_category} shows how the completion rate of trophies changes over time. Among the type (H) games, we see that the completion rates of trophies  generally increase across all the levels. The completion rate of Platinum trophies dramatically increased 4.2\% to 39.2\% from 2008 to 2021. The completion rate of Gold trophies reached even higher at 57.4\% in 2021.
However, does this mean that gamers become more engaged in hunting difficult trophies?  

\begin{figure}[h!]
  \includegraphics[width=\columnwidth]{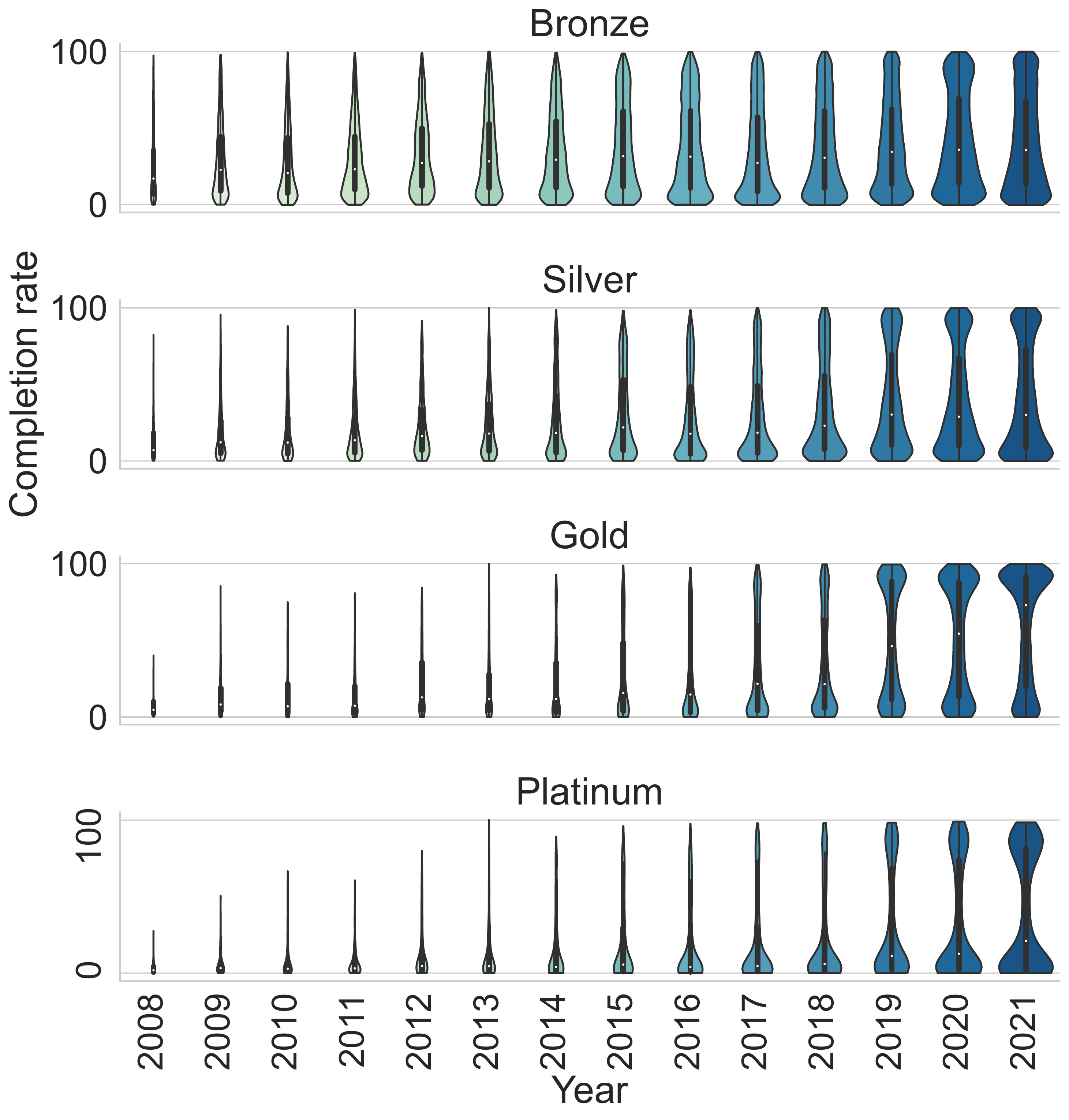}
  \caption{Completion rate of trophies per category over time (type = H)}
  \label{fig:trophy_rarity_per_category_H}
\end{figure}

Figure~\ref{fig:trophy_rarity_per_category_H} can help answer this question. This figure shows the exact distribution of the trophies' completion rates according to their levels. Most of the distributions in the figure are  bi-modal. This means that those trophies that  had been hard to get are still hard to get, but a non-negligible number of easy-to-get trophies have newly appeared. 
This trend has been reported recently~\cite{fastest_platinum}. 
Some users may play games only to obtain  trophies to increase their user-level summed score, to display as a sort of gaming capital~\cite{consalvo2009cheating}, in their PSN profiles. 
To target those users, some games purposefully make their trophies easy to get. This is another type of developers' adaptation to a certain segment of players. 


\subsection{Playtime and Completion Rate}

\begin{figure}[h!]
  \includegraphics[width=\columnwidth]{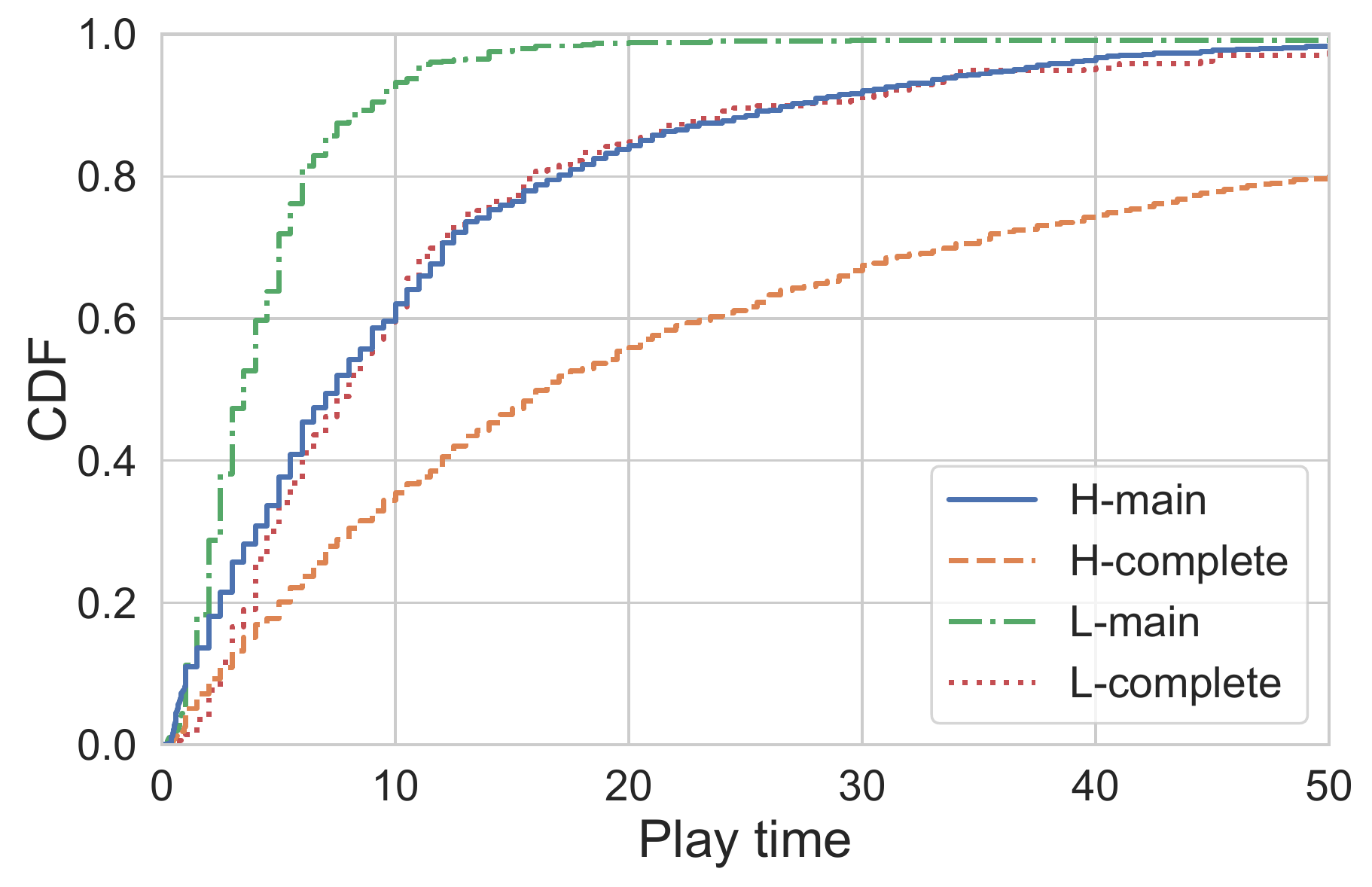}
  \caption{CDF of playtime}
  \label{fig:playtime_CDF}
\end{figure}

Figure~\ref{fig:playtime_CDF} presents the cumulative density function (CDF) of playtime for games of different types (H and L) and different styles (main and complete are mapped into Main Story and Completionist in $\S$\ref{subsec:howlongtobeat}). Like the general belief that a relatively shorter game has around 300 as its summed score of trophies, type (L) games have much shorter playtime (median: 5 hours) than type (H) games (median: 10.5 hours). The average playtime required to complete the game after the main story is 8.6 hours for type (L) games and 23.6 hours for type (H) games. 

\begin{figure}[h!]
  \includegraphics[width=\columnwidth]{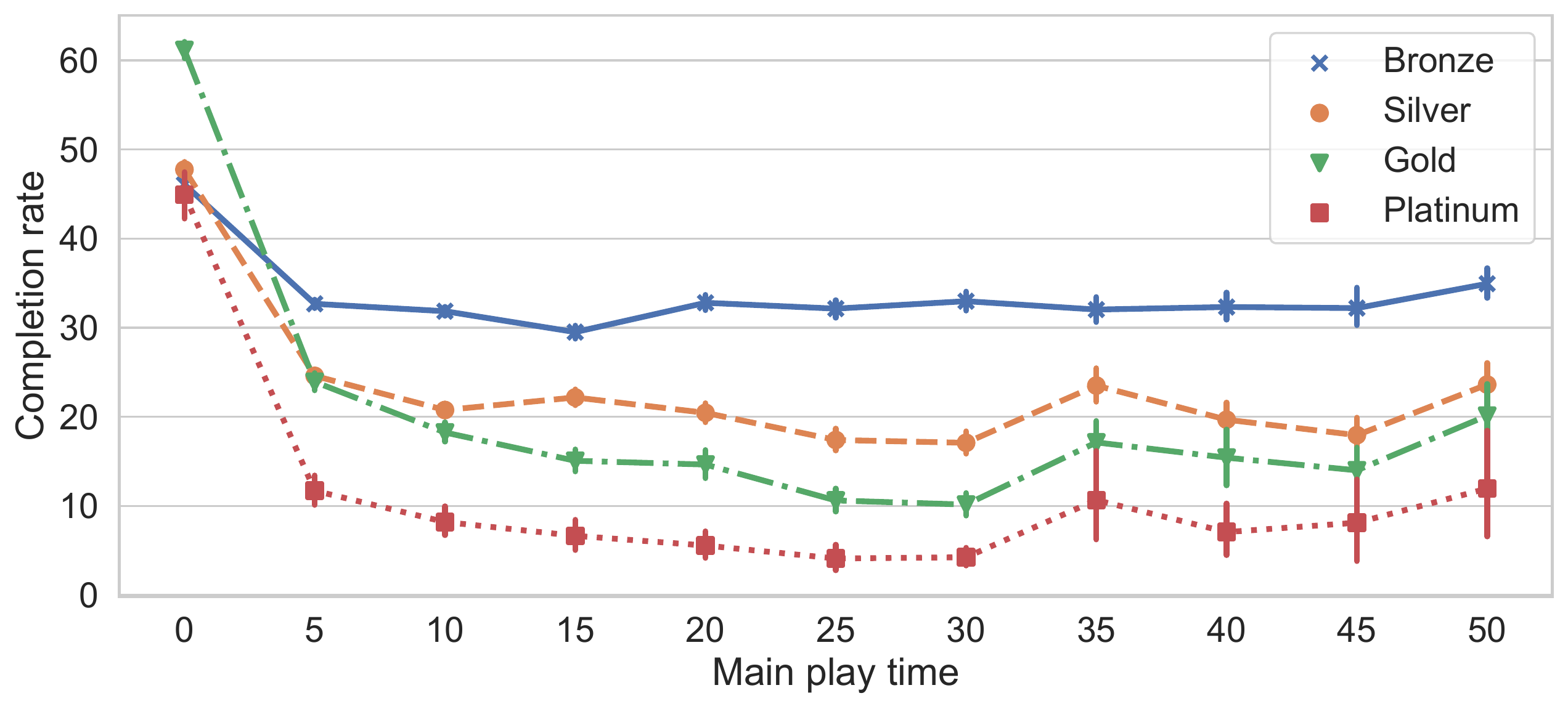}
  \caption{Main playtime and trophy completion rate (type = H)}
  \label{fig:playtime_rarity_H}
\end{figure}

We then move on to how the main playtime correlates with the trophies' completion rates in Figure~\ref{fig:playtime_rarity_H}. Shorter games tend to have trophies with higher completion rates. 
Generally, players can easily get more trophies if the game is short. Also, the existence of easy-to-get trophies reported in Figure~\ref{fig:trophy_rarity_per_category_H} might affect this. 
Another interesting observation is that the  completion rates of Silver, Gold, and Platinum trophies generally decrease until the main playtime is 30 hours, but show some increasing trends  after that. 
The top 5 genres that have longer than 30 hours' main play are Adventure, Role-Playing (RPG), Indie, Strategy, and Tactical, which are considered to fit more hardcore gamers rather than casual gamers. 
Thus, player engagement in hunting trophies might be a driving force to increase the completion rates of trophies for longer ($>$ 30 hours) games. 

\subsection{Genres and Completion Rate}

The genre determines the  gameplay~\cite{heintz2015game}, and the gameplay shapes the trophy design as in Table~\ref{tab:verb_arg1_each_genre}. 
Thus, it is natural to look into how genres are associated with the completion rates of trophies. 

\begin{figure}[h!]
  \includegraphics[width=\columnwidth]{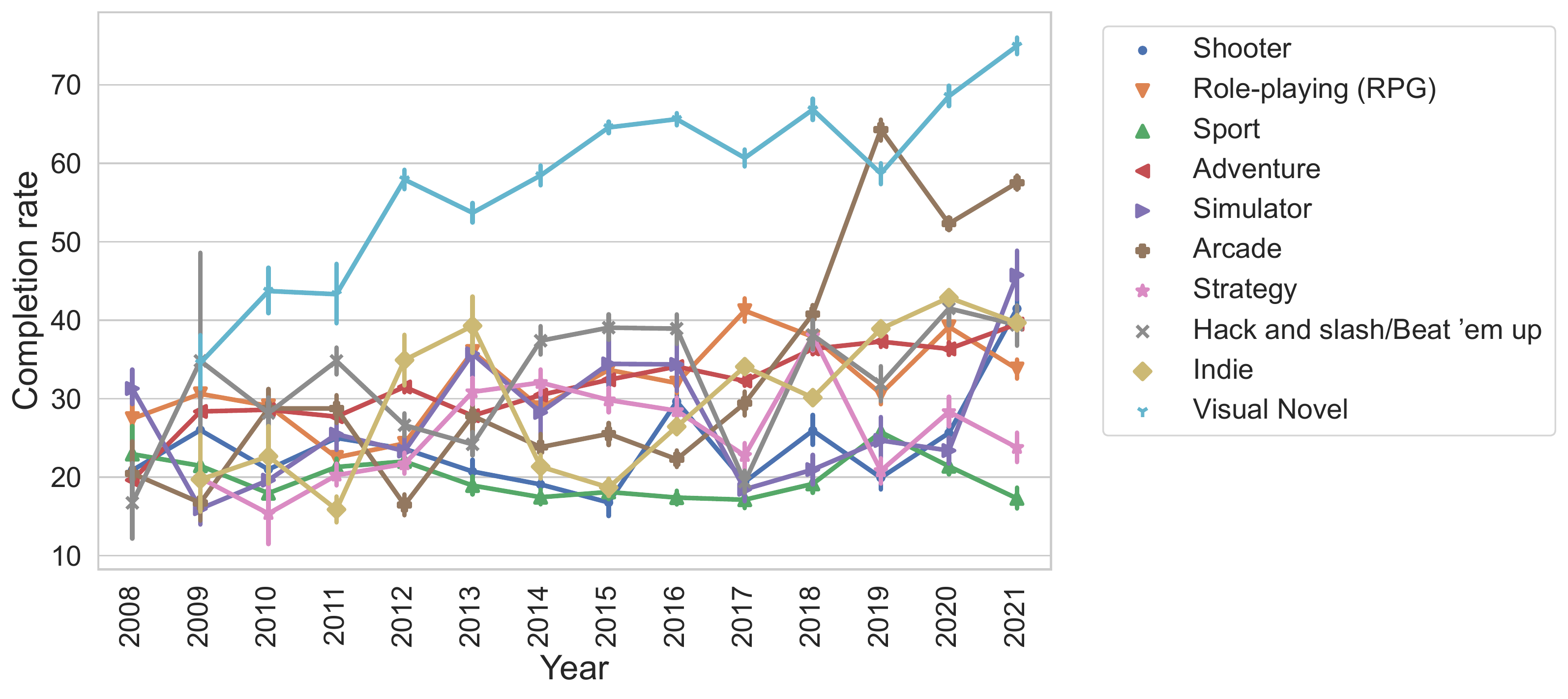}
  \caption{Genres and trophy completion rates  (type = H)}
  \label{fig:genre_rarity_H}
\end{figure}

Figure~\ref{fig:genre_rarity_H} shows the top 10 genres in terms of the number of games and the  completion rates of their trophies over time. While there are some fluctuations in the  completion rates, Visual Novel games consistently show high completion rates. 
These were below 40\% in early years but rapidly increased to 75.0\% in 2021. 
This trend might be interpreted based on the genre characteristics. 
The gameplay in Visual Novel games is minimal. As its name suggests, it is more like reading a novel with graphics and sound; players can click the screen to read the following text or make some narrative choices only. 
Thus, the trophies of Visual Novel games  technically cannot be hard to get because no advanced gameplay is required. Also, as Visual Novel games are niche, players who play them tend to be really into them. As a result, the  players in Visual Novel games are likely to get more trophies than other genres.  

\subsection{Repetitiveness in the Conditions}

As the last analysis to examine players' perspectives, we revisit the trophy condition. As shown in Figure~\ref{fig:N_trophy_cat},   repetitiveness is well embedded in trophy conditions. How does such repetitiveness correlate with the trophy's completion rates? Are there some differences according to the levels of the trophy?

\begin{figure}[h!]
  \includegraphics[width=\columnwidth]{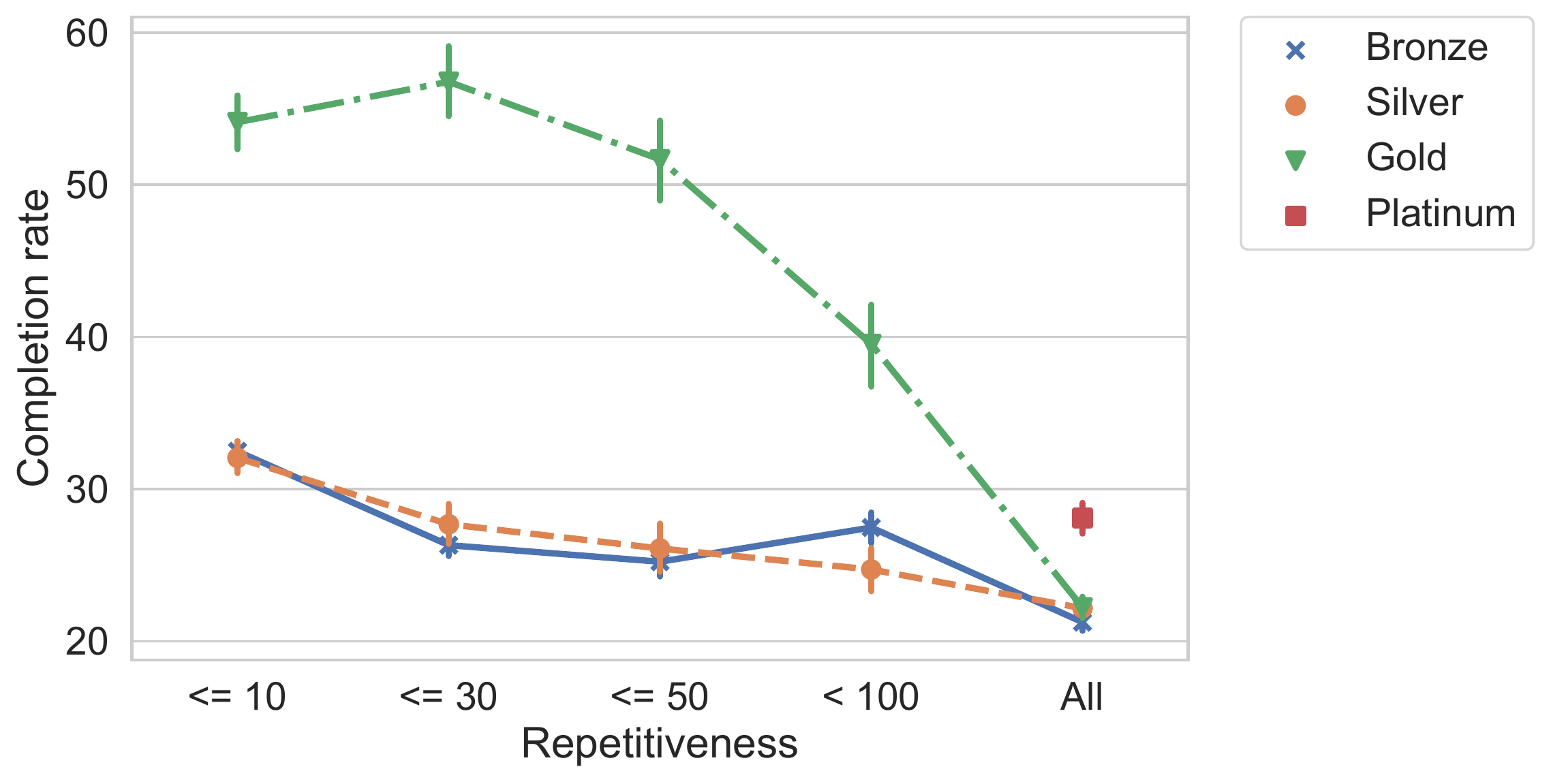}
  \caption{Repetitiveness and trophy completion rate (type=H)}
  \label{fig:N_rarity_pointplot}
\end{figure}

Figure~\ref{fig:N_rarity_pointplot} shows how the  completion rate of a trophy changes with the repetitiveness in its conditions. In addition to the numerical quantifiers, we consider  `any', `every', and `all', which are the top 3 widely used adjectives to represent numbers in trophy conditions. 
We note `100\%' is considered as `All'. We remove Platinum trophies' points except for `All' due to the lack of data points. 
Just as we saw previously in Figure~\ref{fig:trophy_rarity_per_category}, where there was a strong increase in the completion rates of Gold trophies, a high completion rate of Gold trophies can be found here as well. 
Basically, players decrease their likelihood of obtaining trophies when the conditions  become repetitive. However, regarding Gold trophies, the completion rates behave differently. 
Interestingly, the peak for the Gold trophies appears  when the repetitiveness is $<=30$. 
There can be two possible explanations for this.
One is based on the players' behavior. 
To obtain Gold trophies, players might be willing do some repetitive tasks until it becomes not too repetitive ($<= 50$). 
The other is based on the developers' design. As they set the conditions for the same Gold trophies, a task might be easier when $<=$ 30 than when $<=$ 10 even though the repetitiveness is higher. For example, if two conditions are about killing enemies in the same way, `killing 5 enemies' ($<= 10$) could be harder than `killing 15 enemies with a pipe bomb' ($<= 30$) based on the game and game mechanics. 
This uptick behavior is also found in Bronze ($<= 100$). By manual inspection, we find that trophy conditions $<= 100$ are relatively simple tasks, such as `play 100 matches' or `collect 75 logs'. By contrast, trophy conditions $<= 50$ are sometimes harder by specifying a certain item or play style, such as `Shoot 50 objects using Kinesis' or `Perfectly land 50 double backflips'. 

\section{Cross-Platform Games}

So far we examined a single platform, PlayStation Network, among the major gaming platforms. 
As we mentioned in $\S$1, we believe that studying the PlayStation platform can be generalized to other platforms, because 1) the PlayStation Network (PSN) has more than 109 million users who are active monthly as of March 2021~\cite{psn_users}, which means that a huge variety pool of players from casual gamers to hardcore gamers is included in the dataset; 2) PSN supports standard features in major gaming platforms, such as user profiles with a list of `earned' trophies and a timeline for other users' updates. Thus, PSN is a great place to study the impact of common achievement systems; and 3) Many game studios release their games on multiple platforms these days. 
For example, Electronic Arts'  popular football game ``FIFA 2021'' is released on multiple platforms, including PlayStation 4 and 5, Xbox Series X and S, Xbox One, Nintendo Switch, Google Stadia, and Microsoft Windows. Thus, the trophies in games on the PlayStation platforms are just named differently in other platforms (e.g., Achievements in Xbox platforms), implying that how to design trophies is applicable to in-game achievements in general. 
Therefore, we believe that the generalizability of our findings should not be sacrificed even though we studied a single platform.

\begin{figure}[h!]
  \includegraphics[width=\columnwidth]{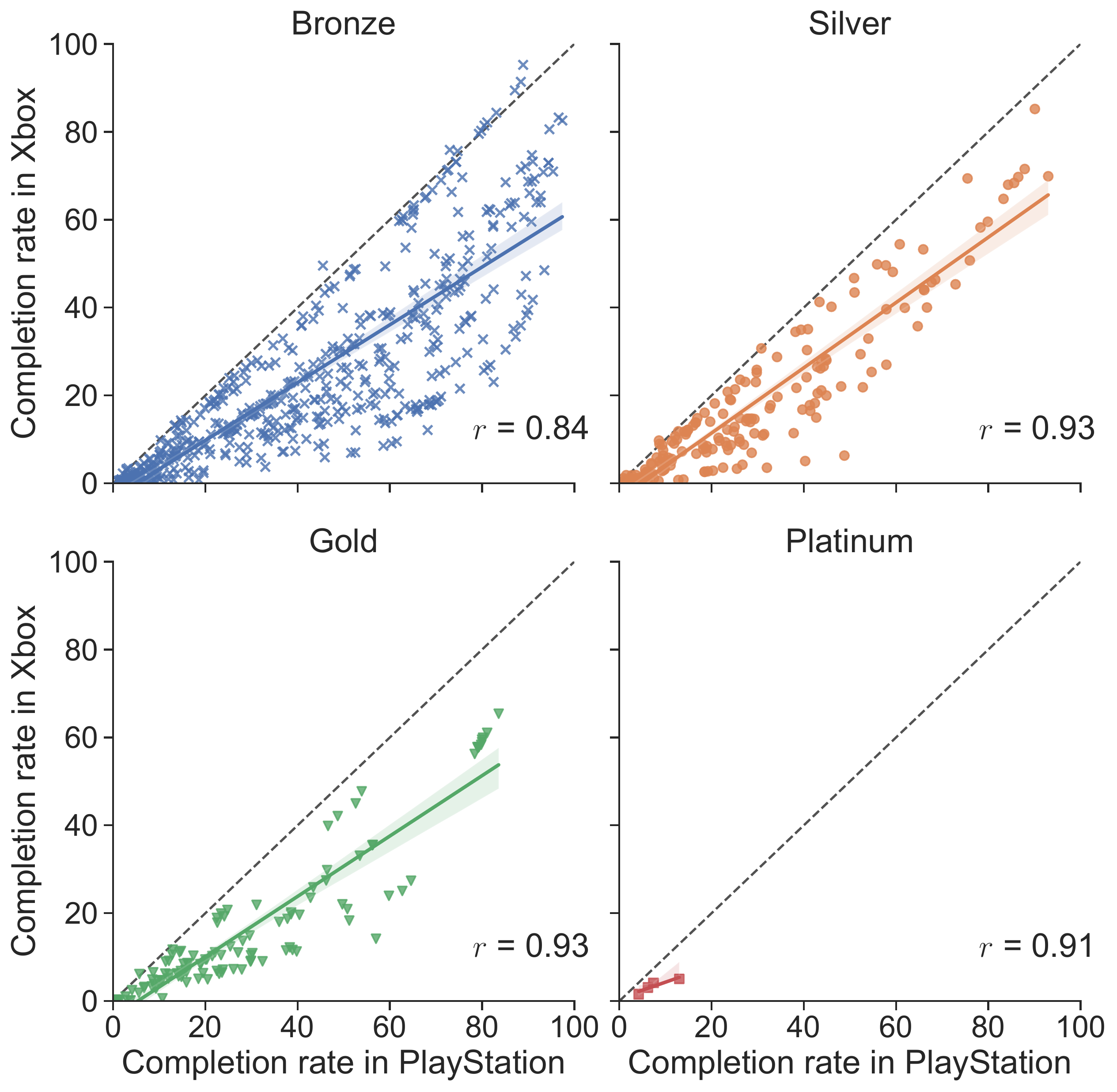}
  \caption{Completion rate of common PlayStation trophies and Xbox achievements (A black dashed line is $y=x$, and a colored-solid line is a linear regression model fit along with a translucent 95\% confidence interval band.)}
  \label{fig:psn_xbox}
\end{figure}

To support our point of view, we conduct a preliminary analysis of cross-platform games. From the top 100 games with the highest review score on OpenCritic in 2021~\cite{opencritic2021best}, we find those released on both PlayStation and Xbox platforms. As Nintendo has not implemented platform-level achievement systems to date, games released for Nintendo Switch are not considered.
Among the top 100 games, we find 45 games are released on the PlayStation platform, and 31 of them (68.9\%) are released on the Xbox platform as well. Among these, we omit 7 DLC (Downloadable content) packs  because they are optional and usually short in length. As a result, we collect trophy designs and completion rates of 24 games between PlayStation 4 and Xbox One. To get information about the achievements (the official name of trophies on the Xbox platform), we use TrueAchievements~\cite{trueachievements}, which is similar to PSNProfiles, but is for the Xbox platform. We find 848 trophies and 836 achievements across these 24 cross-platform games. 

Among them, 828 trophies (or achievements) are common between the two platforms, comprising 97.6\% of the trophies and 99.0\% of the achievements. These extremely high percentages support our ideas about the generalizability of trophy design across  platforms. In other words, our findings about trophy design based on the PlayStation platform can be effectively applied to other platforms as well. 

We also find that the completion rates between the two platform are highly correlated (Pearson $r$=0.87) with each other. This high correlation is consistent also  in   different categories of trophies,  as shown in Figure~\ref{fig:psn_xbox}, supporting the generalizability of gamers' behavior between the platforms. Furthermore, from the figure, it is worth noting that most of the data points are below the $y=x$ line, which means that the completion rate is generally higher in the PlayStation platform than in the Xbox platform. This may imply that the PlayStation platform has  more expert gamers than the Xbox platform, which is similar to our comparison between the dedicated gamer community and the entire gamers in Figure~\ref{fig:trophy_rarity}. Or it might imply that some unique design elements in the PlayStation platform, such as levels of trophies, might elicit more participation. Understanding why PlayStation gamers show higher completion rate of trophies than Xbox gamers will be an interesting future research direction. 


\section{Discussion and Conclusion}

As achievement systems have enormous implications for overall game experience, they have been widely adopted in diverse gaming platforms. 
In this work, we  collected large-scale trophy data from a complete set of games released on Sony PlayStation  platforms. 
We  examined the collected data from both developers' and players' perspectives,  because these are not independent and rather  interact with each other. 

From the developers' perspective, 
we have shown that the number of trophies per game generally decreases over time. Moreover, 
by dissecting the number of trophies per game, we found some interesting internal dynamics. The number of trophies of the lowest level (i.e., Bronze) per game decreases over time, but increases for the higher ones (i.e., Gold).
Both findings support the idea  that the trophy design goal has been changing, with a move toward placing less burdens on players; achievements for harder tasks are more likely to be given, but those for simpler tasks have decreased. 
This change may give positive feedback to players for performing harder tasks and lead to longer playtimes to tackle them. 
In addition, we applied semantic role labeling (SRL) and part-of-speech tagging to the trophy conditions to computationally capture `who did what to whom'~\cite{zhou2015end}. We found that the most frequently appearing Verbs identified by SRL are well aligned with a taxonomy of achievements suggested in previous studies~\cite{montola2009applying}. We suggest that SRL and other natural language processing methods can be further used for large-scale trophy condition analysis, which was not feasible in previous work based on qualitative methods. 

From the players' perspective, we  focused on how many people actually achieve the trophies. By showing a considerable difference in the trophy's completion rates between the entire PSN population and in the  sampled data from a gamer community, we reaffirmed that the data from a gamer community should be carefully analyzed and generalized. 
We then revealed that the completion rates of trophies continuously increase over time.
That might be due to the combined effects of the presence of easy-to-get trophies and the  increasing engagement of gamers in hunting trophies. 
We found that shorter games, in terms of playtime, have trophies with higher completion rates, but interestingly, players are also likely to engage in chasing  trophies in the games longer than 30 hours. 
The trophy completion rate is correlated  with the genre of the games, because a genre determines the gameplay, and the gameplay shapes the trophy design.  Finally, we  investigated the relationship between  repetitiveness in the trophy conditions and the completion rate of the trophy. The relationship does not monotonically decrease or increase; instead, it shows some uptick behavior in the middle, which is probably the result of a trade-off between the  difficulties of an individual task and its repetitions. 

Our preliminary analysis of cross-platform games have shown  the generalizability of our findings across the platforms. More than a half of the top 100 games released on the PlayStation platform are also released on the Xbox platform, and those cross-platform games have highly overlapping trophies (or achievements in the Xbox platform). Furthermore, their completion rates are strongly correlated. 

For future work, we plan to build a prediction model to estimate the difficulty of a trophy based on the dimensions we studied in this work, which have an influence on the completion rate of trophies, including the estimated playtime, genre, trophy condition, and  repetitiveness in the conditions. Such a tool will be helpful for developers to have a better sense of the ideal trophy design before they release a game. 

In summary, we examined how developers have designed achievements to make players more engaged and how players have achieved these  by using large-scale data.
Giving an achievement is a common mechanism to encourage users in many diverse systems beyond just games and appearing under different names, such as badges in gamification. 
We hope that our study will provide insights to researchers on achievement design and user engagement.

\begin{acks}
Kwak gratefully acknowledges the support by D.S. Lee Foundation Fellowship awarded by Singapore Management University.
\end{acks}

\bibliographystyle{ACM-Reference-Format}
\bibliography{websci22-41}


\end{document}